\documentclass{bioinfo}
\copyrightyear{2015} \pubyear{2015}

\usepackage{epstopdf}
\access{Advance Access Publication Date: Day Month Year}
\appnotes{Manuscript Category}

\begin{document}
\firstpage{1}

\subtitle{Genome analysis}

\title[Mechanism for inversion and breakpoint reuse]{Core-genome scaffold comparison reveals the prevalence  that inversion events are associated with pairs of inverted repeats}
\author[Dan Wang \textit{et~al}.]{Dan Wang\,$^{\text{\sfb 1}}$, Shuaicheng Li\,$^{\text{\sfb 1}}$, Fei Guo\,$^{\text{\sfb 3}}$  and Lusheng Wang\,$^{\text{\sfb 1,2,}*}$}
\address{$^{\text{\sf 1}}$Department of Computer Science, City University of Hong Kong, 83 Tat Chee Ave., Hong Kong \\
$^{\text{\sf 2}}$University of Hong Kong Shenzhen Research Institute, Shenzhen Hi-Tech Industrial Park, Nanshan District, Shenzhen, PR China\\
$^{\text{\sf 3}}$  School of Computer Science and Technology, Tianjin University , Tanjin, PR China
}

\corresp{$^\ast$To whom correspondence should be addressed. E-mail: cswangl@cityu.edu.hk}

\history{Received on XXXXX; revised on XXXXX; accepted on XXXXX}

\editor{Associate Editor: XXXXXXX}

\abstract{\textbf{Motivation:} Genome rearrangement plays an important role in evolutionary biology and has profound impacts on phenotype in organisms ranging from microbes to humans. The mechanisms for genome rearrangement events remain unclear. Lots of comparisons have been conducted among different species. To reveal the mechanisms for rearrangement events, comparison of different individuals/strains within the same species (pan-genomes) is more helpful since they are much closer to each other. \\
\textbf{Results:} We study the mechanism for inversion events via core-genome scaffold comparison of different strains within the same species. We focus on two kinds of bacteria, \textit{Pseudomonas aeruginosa} and \textit{Escherichia coli},  and investigate the inversion events among different strains of the same species.  We find an interesting phenomenon that long (larger than 10,000 bp) inversion regions are flanked by a pair of Inverted Repeats (IRs) (with lengths ranging from  385 bp to 27476 bp) which are often Insertion Sequences (ISs). This mechanism can also explain why the breakpoint reuses for inversion events happen. We study the prevalence of the phenomenon and find that it is a major mechanism for inversions. The other observation is that for different rearrangement events such as transposition and inverted block interchange, the two ends of the swapped regions are also associated with repeats so that after the rearrangement operations the two ends of the swapped regions remain unchanged. To our knowledge, this is the first time  such a phenomenon is reported for transposition event.\\
\textbf{Availability and Implementation:} Source codes and examples for our methods are  available at https://drive.google.com/open?id=0B6GLgofcAnevc2hOZEJPYnNieTg \\
\textbf{Contact:} \href{cswangl@cityu.edu.hk}{cswangl@cityu.edu.hk}\\
\textbf{Supplementary information:} Supplemental tables are available at
\\https://drive.google.com/open?id=0B6GLgofcAnevSmRId21CWFR5VGM.}

\maketitle

\section{Introduction}

Comparative genomics studies show that genome rearrangement events often occur between two genomes. Genome rearrangement events play important role
in speciation. The rearrangement operations include deletions, insertions, inversion, transposition, block interchange, translocation, fission and fusion, \emph{etc}.
The mechanisms for those rearrangement events are still unclear. Here we study the mechanism for inversion events
via core-genome scaffold comparison of different strains within the same species.

By comparing two genomes, we can find candidate rearrangement operations. However,
the set of rearrangement operations to transform one genome into the other is not unique in many cases.
Computing the rearrangement operations between two genomes under
different assumptions is an active area, where intensive research have been  conducted \citep{li2006algorithmic}.
It is reported that breakpoints appear more often in repeated regions \citep{lemaitre2009analysis,longo2009distinct}. A summary of the where and wherefore of evolutionary breakpoints is given by \cite{sankoff2009and}. The prevalence of short inversions has been studied \citep{lefebvre2003detection}.  Pevzner and Tesler found extensive breakpoint reuse for inversion events in mammalian evolution
when comparing human and mouse genomic sequences \citep{pevzner2003genome,pevzner2003transforming}.

An interesting problem is to reveal the mechanisms under the rearrangement operations. Many hypothetical mechanisms for the rearrangement operations have been reported \citep{gray2000takes}.
For example, \cite{chen2011genomic} discussed mutational mechanisms for genomic rearrangements. To reveal the mechanisms under the rearrangement operations,
comparison of different individuals/strains within the same species  (pan-genomes) can be more helpful since strains within the
same species  are conserved.

A pan-genome, or supra-genome, describes the full complement of genes
in a clade (typically for species in bacteria and archaea), which can
have large variation in gene content among closely related strains.
Pan-genomes were first studied by Tettelin more than a decade ago \citep{tettelin2005genome}. Several tools have been developed for pan-genome analysis. For example, GET\_HOMOLOGUES \citep{contreras2013get_homologues} is a customizable and detailed pan-genome analysis platform. BLAST atlas \citep{jacobsen2011salmonella} visualizes which genes from the reference genome are present in other genomes.
Mugsy-Annotator \citep{angiuoli2011improving} identifies syntenic orthologs and evaluates annotation quality using multiple whole genome alignments. Characterization of the core and accessory genomes of \textit{Pseudomonas aeruginosa} has been done by \cite{ozer2014characterization}. For pan-genome analysis, genomes from different strains of the same species are decomposed to core blocks (in all the strains), dispensable blocks (in two or more strains) and strain-specific blocks (in one strain only).  Here we extend the pan-genome analysis by comparing the core-genome scaffolds of different strains of the same species.

We study two types of bacteria, \textit{Pseudomonas aeruginosa} and  \textit{Escherichia coli}, and investigate the inversion events among different strains of the same species. We find an interesting phenomenon that long (larger than 10,000 bp) inversion regions are flanked by pairs of Inverted Repeats (IRs) which are often Insertion Sequences (ISs). This mechanism also explains why the breakpoint reuses for inversion events happen. We study the prevalence of the phenomenon and find that it is a major mechanism for inversions.
The other observation is that for different rearrangement events such as transposition and inverted block interchange, the two ends of the swapped regions are also associated with repeats so that after the rearrangement operations the two ends of the swapped regions remain unchanged. To our knowledge, this is the first time  such a phenomenon is reported for transposition event.

\section{Methods}

We develop a pipeline to generate the core-genome blocks, dispensable blocks and strain-specific blocks based on the multiple sequence alignment produced by Mugsy ~(\cite{angiuoli2011mugsy}).

We then develop a computer program to generate the scaffolds of the strains from the core-genome blocks  by repeatedly merging two consecutive blocks appearing in all the strains of the same species. In this way, the number of distinct blocks in the core-genome scaffold is reduced dramatically.  For example, for \textit{Pseudomonas aeruginosa}, before merging, there are 185 blocks in the core genome of the 25 strains. After merging,  the scaffolds contain 69 blocks.

After that, we compute the inversion distance between two scaffolds. Computing the inversion distance between two scaffolds is a very hard and complicated combinatorial problem. Several algorithms have been developed. Due to the difficulty of algorithm design, most of the algorithms only consider inversion events. However, a transposition/block-interchange event can be represented as 3 inversion events, and an inverted transposition/block-interchange event can be represented as 2 inversion events. Therefore, some of the computed inversion events may not be real. There are algorithms dealing with inversion and other rearrangement events such as block interchanges simultaneously. However, the weights for different events are different (again due to the difficulty of algorithm design). Thus, those algorithms still suffer from the problem of outputting inversions that are not real.

Our strategy here is to eliminate some obvious transposition, inverted transposition, block interchange, and inverted block interchange events.

For simplicity, we always assume that  $G_1=+1+2...+n$ is the first input scaffold and $G_2=\pi_1\pi_2\ldots \pi_n$ is  a sign permutation of the $n$ blocks over the set $N=\{1, 2, ..., n\}$ of $n$ distinct blocks, where each integer $i\in N$ appear once in $G_2$ in the form of either $+i$ or $-i$. All the rearrangement operations  are on $G_2$.

A {\it transposition} swaps the order of two consecutive blocks/regions without changing their signs. A transposition $(i,j,k)$ on regions  $\pi_i,\ldots, \pi_{j-1}$  and $\pi_j\ldots \pi_{k-1}$  transforms the sign permutation

$\pi_1\ldots\pi_{i-1}{\bf \pi_i\ldots \pi_{j-1}}{\bf \pi_j\ldots \pi_{k-1}}\pi_k\ldots \pi_n$  

into

$\pi_1\ldots\pi_{i-1}{\bf\pi_j\ldots }
{\bf \pi_{k-1}\pi_i\ldots \pi_{j-1}}\pi_k\ldots \pi_n$.

A transposition    is {\it independent} if it transforms the sign permutation

$\pi_1\ldots\pi_{i-2}\pi_{i-1}{\bf \pi_{i+1} \pi_{i}} \pi_{i+2}\pi_{i+3}\ldots \pi_n$  

into

\begin{figure}[!h]
	\centerline{\includegraphics[width=.49\textwidth]{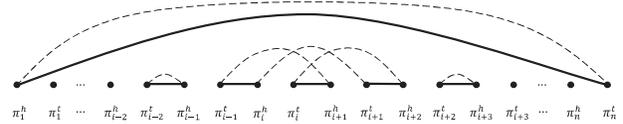}}
	\caption{The breakpoint graph for an independent transposition.}\label{bgraph}
\end{figure}

$\pi_1\ldots\pi_{i-2}\pi_{i-1}{\bf\pi_i \pi_{i+1}}\pi_{i+2} \pi_{i+3}\ldots  \pi_n$,

\noindent where $ \pi_{i-1}{\bf\pi_i \pi_{i+1}}\pi_{i+2}$ is either
$+(q-1)+q+(q+1)+(q+2)$ or $-(q+2)-(q+1)-q-(q-1)$ for $\{q-1, q, q+1, q+2\}\subseteq  N=\{1, 2, ..., n\}$. Though an independent transposition
swaps two consecutive blocks ${\bf \pi_{i+1}}$ and ${\bf  \pi_{i}}$ instead of two regions $\pi_i,\ldots, \pi_{j-1}$  and $\pi_j\ldots \pi_{k-1}$ as in the definition of a general transposition, a pre-process allows us to  merge two consecutive blocks if they are consecutive in both input genomes. Thus, we can still handle some cases for swapping two consecutive regions.
For example, the genome $+1+2+6+7+3+4+5+8$ becomes $+1+2{\bf +4+3}+5$ after merging $+6+7$ (represented as ${\bf +4}$)and  $+3+4+5$ (represented as ${\bf +3}$) and re-number $+8$ as $+5$ in the new representation. An independent transposition can change $+1+2{\bf +4+3}+5$ into $+1+2+3+4+5$.
In terms of breakpoint graph, the two blocks ${\bf \pi_{i+1} \pi_{i}}$ in an independent transposition  is involved in
a 6-edge cycle and after the transformation the 6-edge cycle becomes three 2-edge cycles. In other words, the three breakpoints involved in the 6-edge cycle disappear after the transformation. See Figure \ref{bgraph}.

An {\it  inverted transposition} swaps the order of two consecutive blocks/regions with one of the block's sign changed. An inverted transposition $(i,j,k)$ on regions  $\pi_i,\ldots, \pi_{j-1}$  and $\pi_j\ldots \pi_{k-1}$  transforms the sign permutation $\pi_1\ldots\pi_{i-1}{\bf \pi_i\ldots \pi_{j-1}\pi_j\ldots \pi_{k-1}}\pi_k$
$\ldots \pi_n$  into $\pi_1\ldots\pi_{i-1}{\bf -\pi_{k-1}\ldots} $
${\bf -\pi_{j}\pi_i\ldots \pi_{j-1}}\pi_k\ldots \pi_n$ or  $\pi_1\ldots$  $\pi_{i-1}{\bf \pi_j\ldots \pi_{k-1}-\pi_{j-1}\ldots -\pi_{i}}\pi_k\ldots \pi_n$.

An inverted transposition  is {\it independent} if it transforms the sign permutation
$\pi_1\ldots\pi_{i-2}\pi_{i-1}{\bf -\pi_{i+1} \pi_{i}} \pi_{i+2}\pi_{i+3}\ldots \pi_n$ 

or

$\pi_1\ldots\pi_{i-2}\pi_{i-1}{\bf \pi_{i+1} -\pi_{i}} \pi_{i+2}\pi_{i+3}\ldots \pi_n$ 

into

$\pi_1\ldots\pi_{i-2}\pi_{i-1}{\bf\pi_i \pi_{i+1}}\pi_{i+2} \pi_{i+3}\ldots  \pi_n$,

\noindent where $ \pi_{i-1}{\bf\pi_i \pi_{i+1}}\pi_{i+2}$ is either
$+(q-1)+q+(q+1)+(q+2)$ or $-(q+2)-(q+1)-q-(q-1)$ for $\{q_1, q, q+1, q+2\}\subseteq  N=\{1, 2, ..., n\}$.

A {\it block interchange} swaps the locations of two separated blocks without changing their signs. A block interchange $(i,j,k,l)$ on regions $\pi_i\ldots \pi_j$ and $\pi_k\ldots \pi_l$ transforms

$\pi_1\ldots \pi_{i-1}{\bf \pi_k\ldots \pi_l}\pi_{j+1}\ldots \pi_{k-1} {\bf \pi_i\ldots \pi_j}  \pi_{l+1}\ldots\pi_n$

 into

 $\pi_1\ldots \pi_{i-1}{\bf \pi_i\ldots \pi_j}\pi_{j+1}\ldots \pi_{k-1}{\bf \pi_k\ldots \pi_l}\pi_{l+1}\ldots\pi_n$.

 A block interchange is {\it independent} if it transforms the sign permutation
 
  $\pi_1\ldots \pi_{i-1}{\bf \pi_k}\pi_{i+1}\ldots \pi_{k-1} {\bf \pi_i}  \pi_{k+1}\ldots\pi_n$

 into  
 
\begin{figure}[!h]
	\centerline{\includegraphics[width=.49\textwidth]{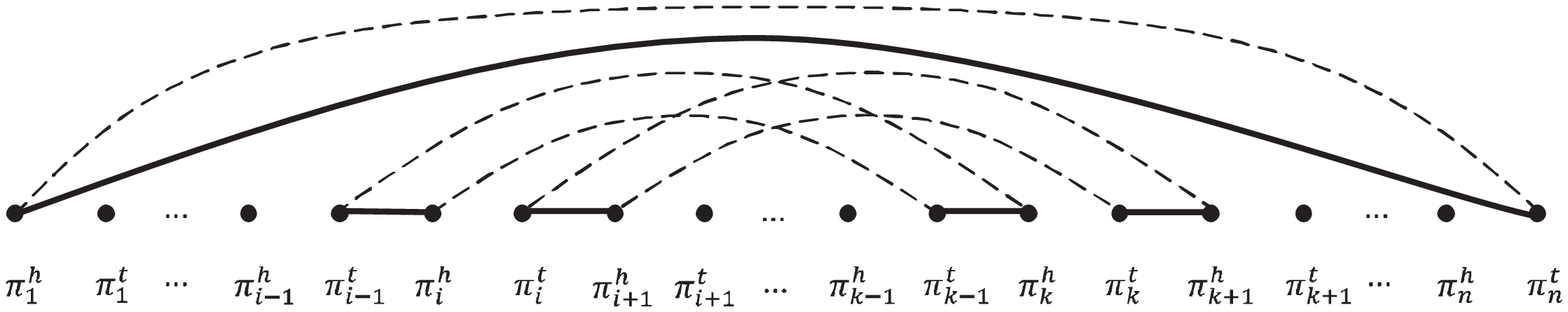}}
	\caption{The breakpoint graph for an independent block interchange.}\label{bbgraph}
\end{figure}

 $\pi_1\ldots  \pi_{i-1}{\bf \pi_i}\pi_{i+1}\ldots \pi_{k-1}{\bf \pi_k}\pi_{k+1}\ldots\pi_n$,

 \noindent where $\pi_{i-1}{\bf \pi_i}\pi_{i+1}$ is either $+q+(q+1)+(q+2)$ or $-(q+2)-(q+1)-q$  and
 $\pi_{k-1} {\bf \pi_k}  \pi_{k+1}$ is either $+p+(p+1)+(p+2)$ or $-(p+2)-(p+1)-p$ for $\{q, q+1, q+2\}\subseteq N$ and
 $\{p, p+1, p+2\}\subseteq N$. Similarly, the two blocks $\pi_k$ and $\pi_i$ are involved in two (interleaving) 4-edge cycles in the breakpoint graph and after the transformation, they become four 2-edge cycles. In other words, there are four breakpoints at the two ends of the two blocks, after
 the transformation, the four breakpoints disappear.  See Figure \ref{bbgraph}.
\ref{bgraph}.

A {\it inverted block interchange} swaps the location of two separated blocks with both signs of the two blocks changed.
A inverted block interchange $(i, j, k, l)$ on  regions $\pi_i\ldots \pi_j$ and $\pi_k\ldots \pi_l$ transforms

$\pi_1\ldots \pi_{i-1}{\bf -\pi_l\ldots -\pi_k}\pi_{j+1}\ldots \pi_{k-1} {\bf -\pi_j\ldots -\pi_i}  \pi_{l+1}\ldots\pi_n$

into

$\pi_1\ldots \pi_{i-1}{\bf \pi_i\ldots \pi_j}\pi_{j+1}\ldots \pi_{k-1}{\bf \pi_k\ldots \pi_l}\pi_{l+1}\ldots\pi_n$.

A inverted block interchange is {\it independent} if it transforms the sign permutation
  $\pi_1\ldots  \pi_{i-1}{\bf -\pi_k}\pi_{i+1}\ldots \pi_{k-1}{\bf -\pi_i}\pi_{k+1}\ldots\pi_n$

 into  $\pi_1\ldots \pi_{i-1}{\bf \pi_i}\pi_{i+1}\ldots \pi_{k-1} {\bf \pi_k}  \pi_{k+1}\ldots\pi_n$,

\noindent
where $\pi_{i-1}{\bf \pi_i}\pi_{i+1}$ is either $+q+(q+1)+(q+2)$ or $-(q+2)-(q+1)-q$  and
 $\pi_{k-1} {\bf \pi_k}  \pi_{k+1}$ is either $+p+(p+1)+(p+2)$ or $-(p+2)-(p+1)-p$ for $\{q, q+1, q+2\}\subseteq N$ and
 $\{p, p+1, p+2\}\subseteq N$.
Again, there are four breakpoints  at the two ends of the two blocks $-\pi _i$ and $-\pi_k$, after
 the transformation, the four breakpoints disappear.

After eliminating independent transposition, inverted transposition,  block interchange and inverted block interchange events, we  use GRIMM-Synteny~  \citep{tesler2002efficient,tesler2002grimm} to compute the inversion distance between pairwise core-genome scaffolds.
We only seriously consider the cases where the rearrangement distance is small.
When the rearrangement  distance is large,  there may be multiple solutions for the inversion history. Thus, in this case, the computed inversion events may not be real.

Finally, we developed a pipeline to compare sequences at the two ends of each inversion region to see whether a pair of inverted repeats exists. Once the inverted repeats are found, the pipeline can also search    all the strains and  mark down its  positions in different strains.

%\enlargethispage{12pt}
\section{Results}

\subsection{\textit{Pseudomonas aeruginosa}}
Complete genome sequences of 25 \textit{Pseudomonas aeruginosa} strains PACS2 (AAQW01000001.1), NCGM1984 (AP014646.1),  NCGM1900 (AP014622.1), NCGM2.S1 (AP012280.1), Carb01\_63 (CP011317.1), SCV20265 (CP006931.1), UCBPP-PA14 (CP000438.1), VRFPA04 (CP008739.2),  DSM\_50071 (CP012001.1), 19BR (AFXJ01000001.1), 213BR (AFXK01000001.1), B136-33 (CP004061.1), PA7 (CP000744.1), LES431 (CP006937.1), PA1 (CP004054.2), YL84 (CP007147.1), LESB58 (FM209186.1), M18 (CP002496.1), RP73 (CP006245.1), DK2 (CP003149.1),   MTB1 (CP006853.1), PAO1 (AE004091.2), F22031 (CP007399.1), PA1R (CP004055.1), and FRD1 (CP010555.1) were downloaded from NCBI GenBank. The details of these 25 \textit{Pseudomonas aeruginosa} strains are listed in Supplemental Table S1. The genome lengths of these strains are between 6.2 mbp (million base pair) and 7.5 mbp.
We used our pipeline to compute the core-genomes and obtained 533 core-blocks with lengths ranging from 58 bp to 83 kbp (kilo base pair) and total lengths ranging from 5.33 mbp to 5.6 mbp (million base pair) which account for 74.8\% - 88.2\% of the strains' genomes.
We then eliminated core blocks with length less than 500 bp and iteratively merged core blocks that were consecutive  for all the 25 strains.  As a result,  69 (merged) blocks were obtained and  the 25 strains led to 8 different scaffolds as shown in Figure \ref{PA_scaf}. The scaffold for each \textit{Pseudomonas aeruginosa} strain is in Supplemental Table S4. For any pair of consecutive blocks in one group,  there must be a different group in which  there is a breakpoint between the  two blocks when comparing the two scaffolds.
\begin{figure}[h]
	\centerline{\includegraphics[width=.48\textwidth]{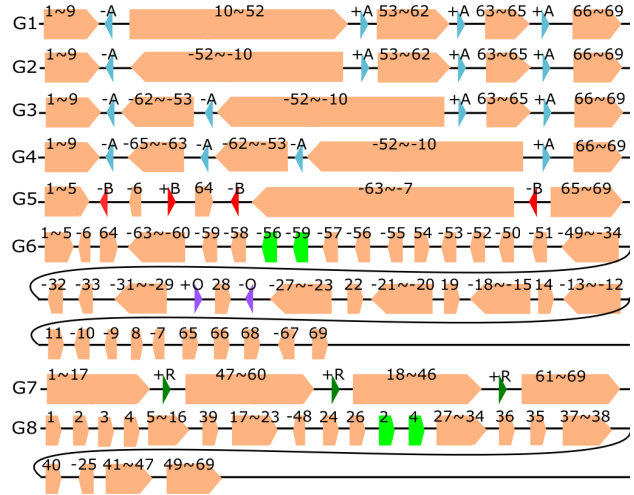}}
	\caption{Eight groups of scaffolds for the 25 \textit{Pseudomonas aeruginosa} strains. Each orange block stands for a merged block which may represent several consecutive core-genome blocks. The numbers above each orange block indicate the included core-genome blocks, for example, 1$\thicksim$5 means the orange block includes five core-genome blocks, which are Blocks  1, 2, 3, 4 and 5. Repeats A, B, O are represented by blue, red and purple triangles respectively. The arrow directions indicate positive/negative strand. }\label{PA_scaf}
\end{figure}

Group 1 contains 13 strains,  which are \textit{Pseudomonas aeruginosa} strains NCGM1984, B136-33, YL84, LESB58, M18, SCV20265, LES431, UCBPP-PA14, DK2, MTB-1, DSM\_50071, Carb01\_63, and F22031. Group 2 contains 6 strains, which are strains RP73, 213BR, PA1, PA1R, 19BR, and PAO1. Groups 3-8 contain 1 strain each and the respective strains are PACS2, FRD1, NCGM2.S1, VRFPA04, NCGM1900, and PA7.

We computed the pairwise inversion distance between scaffolds after eliminating other kinds of independent rearrangement events such as transpositions, inverted-transpositions, block-interchanges, and inverted-block-interchanges. For each of the 8 scaffolds, we chose a scaffold with the minimum inversion distance (after eliminating other independent rearrangement events) to compare. The purpose was to compare two scaffolds with a small number of inversions so that we can observed real inversions between them. From Table \ref{tab:1}, it can be seen that Group 1 is  the closest group to all the other groups except for Group 6. The closest group to Group 6 is Group 5, where the inversion distance
is 7.
\begin{table}[!h]
	\small
	\caption{Shortest inversion distance for each of the 8 groups of \textit{Pseudomonas aeruginosa}.}
	\begin{tabular*}{\linewidth}{@{\extracolsep{\fill}}l|l|l|l|l|l|l}
		%{@{\vrule height 8pt depth3pt  width0pt}l|l|l|l|l|l|l}
		\hline
		$sG$\textsuperscript{a}
		&$cG$\textsuperscript{a}
		&$Inv_d$\textsuperscript{b}
		&$inversion$\textsuperscript{c}
		&$l$\textsuperscript{d}
		&$IR$\textsuperscript{e}
		&$R_d$\textsuperscript{b}\cr
		\hline
		1 & 2 & 1 &  (10,52) & 4.061 & A(2) & 0 \cr
		\hline
		2 & 1 & 1 &  (-52,-10) & 4.061 & A(2)  & 0 \cr
		\hline
		3 & 1 & 1 & (-62,-10) & 4.769  & A(2)  & 0 \cr
		\hline
		4 & 1 & 1 &  (-65,-10) & 5.699 & A(2)  & 0\cr
		\hline
		5 & 1 & 3 &  (-6,-6) & 0.0597   & B(0)  & 0 \cr
		&   &   &    (64,64) & 0.0068    & B(0)    &   \cr
		&   &   &    (-64,-7) & 5.879 & B(0)    &   \cr
		\hline
		6 & 5 & 7 &  (28,28) & 0.0130 & O(0)   & 3 \cr
		&   &   &    (54,54) & 0.0025   & None   &   \cr
		&   &   &    (22,22) & 0.0032   & None   &   \cr
		&   &   &    (19,19) & 0.0031    & None   &   \cr
		&   &   &    (14,14) & 0.0021    & None   &   \cr
		&   &   &    (11,11) & 0.0054    & None   &   \cr
		&   &   &    (8,8) &   0.0074& None   &   \cr
		\hline
		7 & 1 & 0 & None & N/A & N/A  & 1 \cr
		\hline
		8 & 1 & 0 & None & N/A & N/A & 4 \cr
		\hline
	\end{tabular*}
	
	\textsuperscript{a}\footnotesize{Column sG is the source scaffold group, Column cG is the closest scaffold group.}\\
	\textsuperscript{b}\footnotesize{$Inv_d$ indicates the inversion distance between sG and cG after eliminating other independent rearrangement events. $R_d$ indicates the distance of other independent rearrangement events.}\\
	\textsuperscript{c}\footnotesize{The two numbers indicate the starting and ending block of the inversion in the source scaffold (sG). Rearrangement scenario is calculated from the source group to the closest group}\\
	\textsuperscript{d}\footnotesize{$l$ is the length (in Mbp) of inversion of the core-genome segments.}\\
	\textsuperscript{e}\footnotesize{Column $IR$ lists which pair of inverted repeats (A,  B or O) franks the inversion. The numeric code: 0 indicates the respective IR was found only in the source group, 1 indicates the IR was found only in the closest group, 2 indicates the IR was found in both groups.}\\	
	\label{tab:1}
\end{table}
In total,  there are 13 inversion events among the 7 distinct pairs of scaffolds (Table \ref{tab:1}, where pair 1 and 2 appears twice). Among the 13 inversion regions, 7 of them are flanked by a pair of IRs. The remaining 6 inversions with no IRs found at the two ends of the inversion regions  are very short and their  lengths are from 2100 bp to 7400 bp. For each of the first three (Table \ref{tab:1}, rows 1-4)  inversions, the lengths of the inversion regions are more than 4 mbp, and we find a pair of IRs (+A/-A) at the two ends of each of the three long inversion regions. For the pair of Groups 5 and 1, there are three inversions and the lengths of the three inversions in the core-genome are 5.879 mbp, 0.597 mbp, and 6.8 kbp, respectively. Interestingly, we find a repeat $B$ that appears four times in Both Scaffold 1 and Scaffold 5, where $B$ appear as $-B$ once and as $+B$ three times in Scaffold 1. The four occurrences of $B$ form
a pair of IRs at the two ends of each of the 3 inversion regions (See Figure \ref{PA_scaf}).
For  Groups 6 and 5, there exist two independent transpositions and one inverted transposition (See supplement-1). After eliminating the three independent rearrangement events, there are 7 inversions between Groups 6 and 5 which are calculated by GRIMM-Synteny (See Supplement-1) and only one inversion (28,28) is flanked by a pair of IRs (See Table \ref{tab:1}). Note that both $-56$ and $-59$ appear twice in Scaffold 6. We remove the green blocks in Figure \ref{PA_scaf} in our comparison.
   Among these seven inversions, only one inversion (28,28) is longer than 10000bp and flanked by a pair of IRs (+O/-O). Group 1 can be obtained from Group 7 with one independent transposition. A repeat +R appears three times at the ends of the two blocks involved in the transposition.  See  Figure \ref{PA_scaf}). Those occurrences of +R play an important role in the transposition and the details will be discussed  in Section 2.1.2. For Group 8 and 1, there exist two independent transpositions and two independent inverted transpositions (See Supplement-1). After eliminating the four independent rearrangement events, the scaffolds for Group 8 and 1 are actually the same and the inversion distance between them is zero.   Again,   both Blocks 2 and Block 4 appear twice in  Group 8. (The physical positions of all the copies of Blocks 2 and 4 in  Group 8 are in Supplemental Table S5h). We remove the green blocks in Figure \ref{PA_scaf} in our comparison.

For the first inversion between Group 1 and 2, there are 13 strains in Group 1 and 6 strains in Group 2. All the strains in Group 1 and Group 2 contain Repeat $+A$ and $-A$ as shown in Figure \ref{PA_scaf}. The physical positions as well as the lengths of the repeats differ slightly in different strains. See Supplemental Table S5a.
Thus, the inversion (from Blocks 10 to 52) between Scaffold 1 and Scaffold 2  (row 1 in Table \ref{tab:1}) is found between the $13\times 6$ pairs of strains in these two groups.
For the remaining inversions listed in Table \ref{tab:1}, the physical positions,  the lengths of repeats and core-genome blocks (at
the two ends of an inversion) in different strains are given in Supplemental Tables S5b-e.

In summary,  three different pairs of IRs are found and we use +A/-A, +B/-B and +O/-O to differentiate these three pairs. We also find three copies of +R in comparison of Groups 1 and 7. The locations of these repeats in the scaffolds are shown in Figure \ref{PA_scaf}.  The lengths (in bp), gene products and protein IDs (in NCBI Protein database) of these  repeats are listed in Supplemental Table S9.
%Among these 4 repeats, +B/-B contains a gene which encodes transposase.
\subsubsection{Breakpoint reuse}
\begin{figure}[!h]
	\centerline{\includegraphics[width=.49\textwidth]{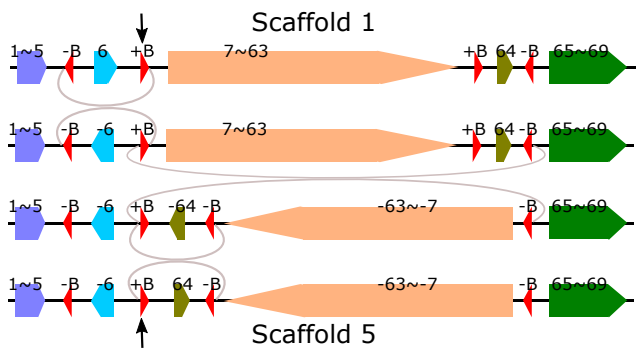}}
	\caption{Three inversion steps from scaffold  1 to scaffold 5. The breakpoint between -6 and 64  in Scaffold 5 is used three times. See the black arrow. }\label{PAG1G5}
\end{figure}

The three inversion steps from Scaffold 1 to 5 are shown in Figure \ref{PAG1G5}, where it can be seen that there is a +B and three -Bs in Scaffold 5.  The three inversion events are -B-6+B to -B6+B, +B7$\thicksim$64-B to
+B-64$\thicksim$-7-B and +B-64-B to +B64-B and the breakpoint  the black arrow points at in Figure \ref{PAG1G5} is used three times.

Here +B plays a crucial role in the three inversions and is used three times, each time +B and -B form a pair of inverted repeats at the two ends of the inversion regions.
Now let us  have a close look at +B (of length 820 bp), we can see that for the first inversion (-B-6+B to -B6+B), the real cutting points (breakpoints) are at the left end of -B and the right end of +B, while for the other two inversions (+B7$\thicksim$64-B to
+B-64$\thicksim$-7-B and +B-64-B to +B+64-B), the real cutting points (breakpoints) are at the left end of +B and the right end of -B. Here the real cutting point does not seem to be important and the  repetitive element B should be viewed as the breakpoint.

Another interesting finding is that for Groups 1, 2, 3 and  4, each scaffold contains a -A and three +As. (See Figure \ref{PA_scaf}.) Theoretically, this -A can be reused three times with each of the three +As. However, we did not observe such three breakpoint reuses in a single pairwise scaffold comparison. But it has been observed that this -A, along with each of the three +As, mediate three different inversion events which occur between Group 1 and Group 2, Group 1 and Group 3, and Group 1 and Group 4,  respectively (Table \ref{tab:1}, row 2-4).

\subsubsection{Transposition}
\begin{figure}[h]
	\centerline{\includegraphics[width=.48\textwidth]{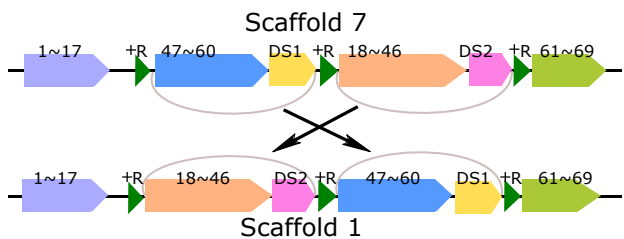}}
	\caption{ Both Scaffolds 1 and  7 contain four merged core  blocks (1$\thicksim$17), (18$\thicksim$46), (47$\thicksim$60), and  (61$\thicksim$69). Moreover, both Scaffolds 1 and  7 contain another two non-core blocks DS1 and DS2, where the occurrences of DS1 and DS2 in both scaffolds are 100\% identical. There are three occurrences of a repeat +R in both scaffolds. }\label{PA_G1G7}
\end{figure}

 Figure \ref{PA_G1G7} gives the detailed scaffolds for Groups 1 and 7. Both Scaffolds 1 and  7 contain four merged core  blocks (1$\thicksim$17), (18$\thicksim$46), (47$\thicksim$60), and  (61$\thicksim$69). Moreover, both Scaffolds 1 and  7 contain another two non-core blocks DS1 and DS2, where the occurrences of DS1 and DS2 in both scaffolds are 100\% identical. Besides, there are three occurrences of a repeat +R in both scaffolds. It can be seen that by swapping 47$\thicksim$60 and DS1 with 18$\thicksim$46 and DS2, Scaffold 7 is transferred into Scaffold 1.  The most interesting finding is the  three occurrences of +R  located at the three breakpoints of the transposition.
We believe that this three occurrences of +R play an important role in this transposition event because the repeat +R can make sure the two ends of the two swapped regions remain unchanged before and after the transposition. This is similar to the mechanism that inversion regions are franked by a pair of IRs, where after the inversion the two ends of the inversion region remain the same. For reference, the physical positions of the three +Rs, DS1, DS2 and Blocks 47, 60, 18 and 46 in the chromosomes of Group 7 and 1 are listed in Supplemental Table S5f.
\subsection{\textit{Escherichia coli}}
We selected 31 \textit{Escherichia coli} strains (identification number (id) 1 to 31) with complete sequences from  17 genome families at NCBI's GenBank. These 31 strains are SE15 (AP009378.1), IAI39 (CU928164.2), EC4115 (CP001164.1), CFT073 (AE014075.1), CE10 (CP003034.1), O103:H2 str. 12009 (AP010958.1), C227-11 (CP011331.1), 536 (CP000247.1), K-12 substr. MG1655 (U00096.3), ST2747 (CP007392.1), NA114 (CP002797.2), 042 (FN554766.1), O111:H- str. 11128 (AP010960.1), O145:H28 str. RM13514 (CP006027.1), O104:H4 str. 2011C-3493 (CP003289.1), SE11 (AP009240.1), SS52 (CP010304.1), APEC O78 (CP004009.1), SMS-3-5 (CP000970.1), DH1Ec095 (CP012125.1), 1303 (CP009166.1), O157:H7 str. Sakai (BA000007.2), 55989 (CU928145.2), B str. REL606 (CP000819.1), O83:H1 str. NRG 857C (CP001855.1), UMN026 (CU928163.2), PCN033 (CP006632.1), 789 (CP010315.1), O127:H6 str. E2348/69 (FM180568.1), P12b (CP002291.1), and ED1a (CU928162.2). The detailed information of these 31 strains is listed in Supplemental Table S2. The genome lengths of these strains are between 4614223 bp and 5585613 bp. Our pipeline found 344 core blocks. The lengths of these core blocks range  from 45 bp to 72931 bp and the total core-genome lengths in different strains range from 4006932 bp to 4246034 bp which account for 74.07\%-88.42\% of the strains' genomes. After eliminating core-blocks
with length less than 500 bp and repeatedly merge two consecutive core-blocks (that are  consecutive  for all
the 31 strains), we obtained 49 (merged) blocks and the 31 strains formed 9 groups of scaffolds (G1-G9 as shown in Figure \ref{Ecoli_scaf}). The scaffold for each of the 31 \textit{Escherichia coli} strain is given in Supplemental Table S6.

Group 1 contains 21 strains which are \textit{Escherichia coli} strains EC4115, CE10, C227-11, K-12 substr. MG1655, ST2747, 042
O104:H4 str. 2011C-3493, SE11, SS52, APEC O78, DH1Ec095, 1303, O157:H7 str. Sakai, 55989, B str. REL606, O83:H1 str. NRG 857C, UMN026, PCN033, 789, O127:H6 str. E2348/69, and ED1a. Group 2 contains 3 strains,  SE15, CFT073 and 536. Groups 3-9 contain 1 strain each and the respective strains are O145:H28 str. RM13514, SMS-3-5, P12b, IAI39, O103:H2 str. 12009, NA114, and O111:H- str. 11128.

After computing pairwise inversion distance among the 9 scaffolds, we selected a scaffold with minimum inversion distance for each of the 9 scaffolds as shown in Table \ref{tab:4} for comparison. From Table \ref{tab:4}, it can be seen that Group 1 is the closest group to all the other 8 groups with inversion distances ranging from 0 to 4. The closest group to Group 1 is Group 2, where the sign of Block 24  is different.
\begin{figure}[h!]
	\centerline{\includegraphics[width=.49\textwidth]{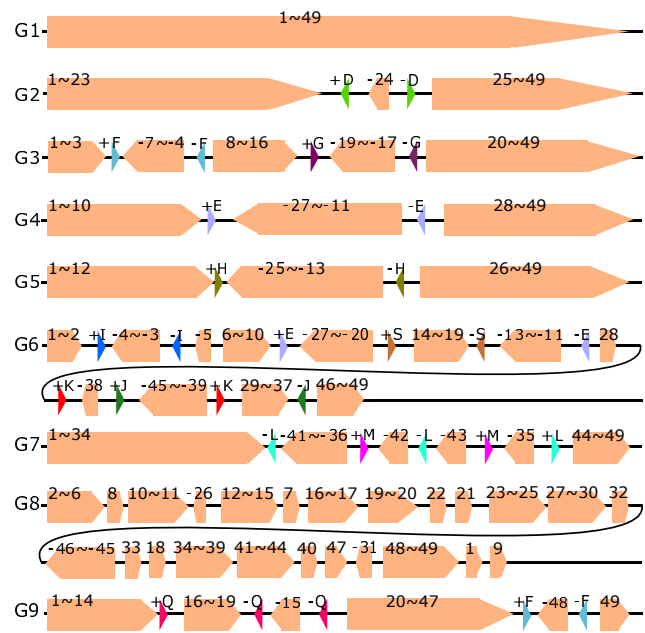}}
	\caption{Nine groups of scaffolds for the 31 \textit{Escherichia coli} strains}\label{Ecoli_scaf}
\end{figure}
\begin{table}[!h]
	\small
	\caption{Shortest inversion distannce for each of the 9 groups of \textit{Escherichia coli}.}\label{tab:4}
	\begin{tabular*}{\linewidth}{@{\extracolsep{\fill}}l|l|l|l|l|l|l}
		%{@{\vrule height 8pt depth3pt  width0pt}l|l|l|l|l|l|l}
		\hline
		$sG$
		&$cG$
		& $Inv_d$
		& $inversion$
		& \multicolumn{1}{c|}{$l$}
		& \multicolumn{1}{c|}{$IR$}
		& $R_d$\cr
		\hline
		1 & 2 & 1 &  (24,24) & 0.0041 & D(1)\textsuperscript{a} & 0 \cr
		\hline
		2 & 1 & 1 &  (-24,-24) & 0.0041 & D(0)\textsuperscript{a}  & 0 \cr
		\hline
		3 & 1 & 2 & (-7,-4) & 0.2763  & F(0)  & 0 \cr
		&   &   &    (-19,-17) & 0.1940   & G(0)\textsuperscript{b}  &   \cr
		\hline
		4 & 1 & 1 &  (-27,-11) & 1.402 &E(0)  & 0\cr
		\hline
		5 & 1 & 1 &  (-25,-13) & 1.111   & H(0)  & 0 \cr
		\hline
		6 & 1 & 4 &  (-4,-3) & 0.2706 & I(0)   & 1 \cr
		&   &   &  (-5,-5) & 0.0075  &  --   &  \cr
		&   &   &  (-45,37) & 1.4108  & J(0) &  \cr
		&   &   &  (-38,-29)  & 1.2756\textsuperscript{c}  & K(0) & \cr
		\hline
		7 & 1 & 3 & (-43,-35) & 0.0642 & L(0)  & 0 \cr
		&   &   &  (-42,35) & 0.1055\textsuperscript{d}& M(0)  &  \cr
		&   &   &  (-41,-35) & 0.3944\textsuperscript{e}&  L(0)   &  \cr
		\hline
		8 & 1 & 4 & (See Supplement-1) & N/A & N/A  & 6 \cr
		\hline
		9 & 1 & 1 & (48,48)  & 0.0651 & F(0) & 1 \cr
		\hline
	\end{tabular*}
	
	\textsuperscript{a}\footnotesize{In Group 1, only Strain SE15 has +D/-D at the ends of 24}\\
	\textsuperscript{b}\footnotesize{In Group 2, only Strain O157:H7 str. Sakai has +G/-G at the ends of (-19,-17)}\\
	\textsuperscript{c}\footnotesize{$l$=length of Block 38 + length from Block  29 to Block 37 in Group 6.}\\
	\textsuperscript{d}\footnotesize{$l$=length of Block 42 + length of Block 35 in Group 7.}\\
	\textsuperscript{e}\footnotesize{$l$=length from Block 41 to Block 36 + length of Block  35 in Group 7.}\\	
	\label{tab:4}
\end{table}
In total, there are 17 inversion events among the 8 distinct pairs in Table \ref{tab:4} (the pair of Group 1 and Group 2 appears twice) and the inversion region lengths varies from 0.0075 mbp to 1.402 mbp. (See Table \ref{tab:4}.)
Among the 17 inversion regions, 12 of them  are found to be flanked by a pair of inverted repeats in the strains of the source groups. For  inversion (-5,5)  between Group 1 and Group 6 (row 6 in Table \ref{tab:4}) and the four inversions between Group 1 and 8,  no pairs of inverted repeats are found at the two ends of the block. The length of  inversion (-5,5) (Row 6 in Table  \ref{tab:4}) is 
short (7.5 kbp). The four computed inversions between Group 1 and 8 may not be true since there are another 6 other rearrangement events between the two scaffolds (Row 8 in Table  \ref{tab:4}).
   For Groups 6 and 1, the rearrangement distance is five (one independent inverted block interchange and a sequence of four inversions). See Table \ref{tab:4}. At the breakpoints of this inverted block interchange, we also find IRs and we will discuss it in Section 2.2.2. For Group 8 and 1, after eliminating six independent transpositions, there exists a sequence of four inversions (See Supplement-1). Only one of these four inversions is flanked by a pair of IRs. We observe that  there are seven copies of Block 45 in  Group 8 and we used the -45 next to -46 for comparison.
The distance between Group 1 and Group 8 is big (6 transpositions + 4 inversions) and thus our  predicted  rearrangement history between Group 1 and Group 8 may not be correct. (Again, for reference, the physical positions of these seven copies of Block 45 in the chromosome of Group 8 are in Supplemental Table S7i.)
 To obtain Group 1 from Group  9, an independent inverted transposition and an inversion (Block -48 in Scaffold 9) are required.  (See Table \ref{tab:4}). The inverted region (Block -48) is flanked by a pair of IRs (+F/-F) in the  Group 9. (See Figure \ref{Ecoli_scaf}.) In addition, we find that this inverted transposition event is also associated with repeats and we will discuss this  in Section 2.2.3.

For all the inversions listed in Table \ref{tab:4}, the physical positions, the lengths of repeats and core-genome blocks (at the two ends of  inversions) in different strains are given in Supplemental Table S7a-g.

We find a total of 12 different types of pairs of inverted repeats and use letters from +D/-D to +M/-M, +S/-S and +Q/-Q to label and differentiate these 12 pairs of IRs. The locations of these IRs in the scaffolds are shown in Figure \ref{Ecoli_scaf}. The lengths (in bp), gene products and protein IDs (in NCBI Protein database) of these 12 IRs are listed in Supplemental Table S8. We note that 7 of these 12 pairs of IRs contain genes which encode transposase.

\subsubsection{Breakpoint reuse}
\begin{figure}[h]
	\centerline{\includegraphics[width=.49\textwidth]{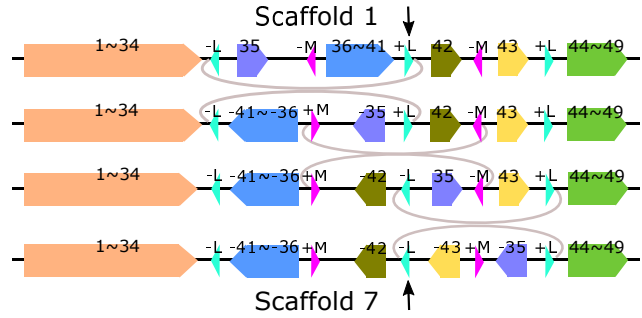}}
	\caption{Three inversions between Scaffolds  1 and  7. The breakpoint between 41 and 42 in Scaffold 1 is used twice. See the black arrow. }
	\label{ECG1G7}
\end{figure}

The three inversion steps from Scaffolds 1 to  7 are illustrated in Figure \ref{ECG1G7}. From Figure \ref{ECG1G7}, it can be seen that The breakpoint between 41 and 42 in Scaffold 1 is used twice. The corresponding inversion regions are flanked by -L and +L.

It is worth  pointing out that the two +Ms in Scaffold 1 form a pair of directed repeats (DRs). After  inversion (35,-41), the pair of directed repeats (DRs) of M becomes a pair of inverted repeats. This means that a pair of DRs has the potential to mediate inversions.

\subsubsection{Inverted Block Interchange}
\begin{figure}[h]
		\centerline{\includegraphics[width=.49\textwidth]{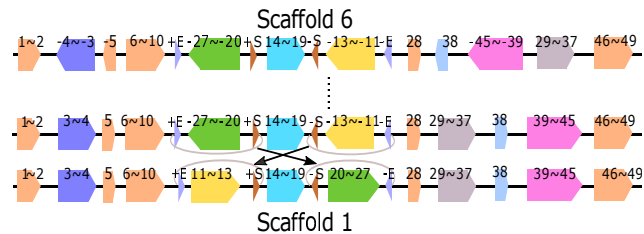}}
		\caption{Inverted block interchange of Region -27$\thicksim$-20 and Region -13$\thicksim$-11 between Scaffolds 6 and 1. +E/-E and +S/-S are two pairs of IRs. The steps from  Scaffold 6 to the middle scaffold are omitted.}
		\label{ECG1G61}
	\end{figure}

We find an inverted block interchange between Scaffold 6 and 1 and we use Figure \ref{ECG1G61} to illustrate.  In Figure \ref{ECG1G61}, Region +E-27$\thicksim$-20+S and --S13$\thicksim$-11-E in Scaffold 6 are inversely interchanged with each other to obtained Scaffold 1.
The existence of two pairs of IRs (+E/-E and +S/-S) makes sure the two ends of the swapped blocks remain unchanged after the inverted block interchange event. The physical positions of +E/-E, +S/-S and Blocks 27, 20, 13 and 11 in  Groups 6 and 1 are listed in Supplemental Table S7h.

\begin{figure}[h]
		\centerline{\includegraphics[width=.49\textwidth]{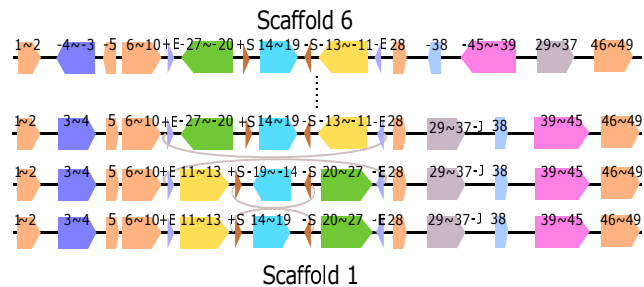}}
		\caption{Two inversions which can replace the inverted block interchange of Regions -27$\thicksim$-20 and -13$\thicksim$-11 between Scaffold 6 and 1. The first inversion is flanked by +E and -E and the second inversion is flanked by +S and -S. The steps from the Scaffold 6 to its next scaffold are omitted. }
		\label{ECG1G62}
	\end{figure}

The other explanation is that an inverted block interchange can be replaced by two inversions. Figure \ref{ECG1G62} shows the two inversions which can replace the inverted block interchange of Blocks -27$\thicksim$-20 and Block -13$\thicksim$-11.  Each of these two inversions is flanked by a pair of IRs (See Figure \ref{ECG1G62}).
	
\subsubsection{Inverted Transposition}
\begin{figure}[h]
		\centerline{\includegraphics[width=.49\textwidth]{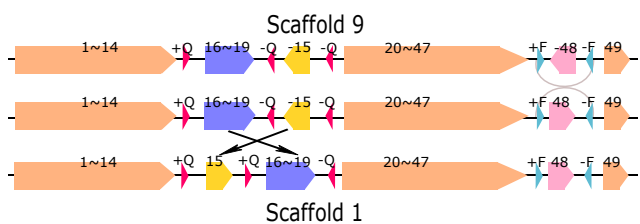}}
		\caption{Inverted transposition of Region 16$\thicksim$19 and Block 15 between Scaffold 9 and 1. There are three occurrences of Repeat Q with different signs. From Scaffold 9 to the next scaffold, there is an inversion of Block 48 which are flanked by +F and -F.}
		\label{ECG1G91}
	\end{figure}

Figure \ref{ECG1G91} shows the inverted transposition from Scaffolds 9 to 1: Block -15 and region 16$\thicksim$18 in Scaffold 9 are swapped with each other  with the sign of Block -15  changed.  Block -15 is flanked by a pair of directed repeats (DRs) (-Q,-Q) and Region 16$\thicksim$18 is flanked by a pair of IRs (+Q,-Q) in Scaffold 9. These three occurrences of Repeat Q can make the ends of Block -15 and Block 16-18 remain unchanged  after the inverted transposition (with the sign of Block -15 changed). The physical positions of the three copies of Repeat Q and Blocks 15, 16 and 18 in the chromosomes of Group 9 and 1 are listed in Supplemental Table S7j.

\begin{figure}[h]
		\centerline{\includegraphics[width=.49\textwidth]{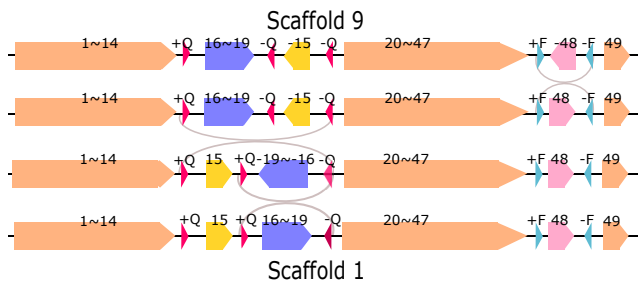}}
		\caption{Two inversions which can replace the inverted transposition of Region 16$\thicksim$19 and Block 15 between Scaffold 9 and 1. Both of the two inversions are flanked by a pair of IRs (+Q/-Q). From Scaffold 9 to the next scaffold, there is an inversion of Block 48 which are flanked by +F and -F.}
		\label{ECG1G92}
	\end{figure}

The other explanation is that the inverted transposition can be replaced by two inversions: the first inversion is from Blocks 16 to -15 and the second inversion is from Blocks -19 to -16 (See Figure \ref{ECG1G92}). Both of these two inversions are flanked by a pair of IRs (+Q/-Q).
	
\section{Discussions}

Breakpoint reuses for inversion event was first reported by \cite{pevzner2003human} when comparing
human and mouse genomic sequences.
Their  analysis shows  that at least 245 rearrangements of 281
synteny blocks occurred between  human and
mouse genome. It is estimated that any human \& mouse
rearrangement scenario requires at least 190 breakpoint reuses \citep{pevzner2003human}.
Their analysis on X chromosome between  human and mouse illustrates that there are two different most parsimonious scenarios.
Both contain  three (different) breakpoint reuses. \cite{sankoff2005chromosomal} show that breakpoint reuse is very sensitive to the proportion of
blocks excluded. \cite{attie2011rise} also show that the inferred breakpoint reuse rate depends on synteny block resolution
in human-mouse genome comparisons.
Statistics analyzes showed that breakpoints are often associated with repetitive elements
and the density of breakpoints in small intergenes appears significantly higher than in gene deserts \citep{longo2009distinct,sankoff2009and}.
However, the mechanism for breakpoint reuse is not clear. Here our observation that long inversions are flanked  by a pair of inverted repetitive elements can
clearly explain why breakpoint reuse happens for inversions. The comparative results for the two  different kinds of bacteria also illustrate the prevalence of this phenomenon.
Recently, breakpoint reuse for inversions has been reported in \textit{Drosophila} genus \citep{puerma2014characterization,orengo2015molecular}
as well as \textit{Saccharomyces pastorianus} \citep{hewitt2014sequencing}.
\cite{rajaraman2013fpsac}  suggested that rearrangements could be driven by the ISs and the positions of the inversion breakpoints in their study were also highly
correlated with IS : 76 of
the 118 mapped breakpoints were close (<1000 nt distant) to some predicted IS, whereas this number drops to 39 for uniformly sampled random coordinates (P-value <$ 10^{-3} $).
\cite{darmon2014bacterial} reviewed many examples of  prokaryotic genomic rearrangements which were induced by natural transposable elements and pointed out that   recombination between IRs can result in an inversion of the internal DNA sequence.
The association between IR and  genome rearrangement breakpoints was also reported in previous studies on mammals and drosophila genomes \citep{thomas2011genome,ranz2007principles,bailey2004hotspots,armengol2003enrichment,samonte2002segmental}.  Accounting for this phenomenon in order to reduce the space of optimal inversion scenarios was explored. \cite{armengol2003enrichment} observed that nine primary regions involved in human genomic disorders which show changes in the order or the orientation of mouse/human synteny
segments were often flanked by segmental duplications in the human sequence. They also found that $53\%$  of
all evolutionary rearrangement breakpoints associate with segmental duplications, as compared with $18\%$ expected in a random location of breaks along the chromosome (P < $ 10^{-4} $). \cite{ranz2007principles} analyzed the breakpoint regions of the 29 inversions that differentiate the chromosomes of \textit{Drosophila
melanogaster} and two closely related species,\textit{ D. simulans} and \textit{D. yakuba}, and reconstructed the molecular events that
underlie their origin. Experimental and computational analysis revealed that the breakpoint regions of 59\% of the
inversions (17/29) are associated with inverted duplications of genes or other non-repetitive sequences.
They also for the first time  reconstruct the reuse of a breakpoint region in
Diptera \citep{ranz2007principles}.

\section*{Dataset}
\textbf{\textit{Pseudomonas aeruginosa} data set.} On 10 June 2015, We downloaded all of the publicly available \textit{Pseudomonas aeruginosa} complete genomes from GenBank at the NCBI. The detailed information of the 25 Pseudomonas aeruginosa strains is listed in Supplemental Table S1.\\
\textbf{\textit{Escherichia coli} data set.} At NCBI's GenBank, there are total 150 complete genomes and 33 genome groups. 21 genome groups have complete genomes for \textit{Escherichia coli}. We downloaded complete genomes of 31 \textit{Escherichia coli} strains from 17 different genome groups and the detailed information of our dataset is in Supplemental Table S2.\\
%\textbf{\textit{Shewanella} data set.} All the 24 strains with Complete genomes in \textit{Shewanella} Genus were downloaded from NCBI's %GenBank. The detailed information of these 24 strains is listed in Supplemental Table S3.
\section*{Funding}
This work is supported by a National Science Foundation of China (NSFC
61373048) and a grant from the Research Grants Council of the Hong Kong Special
Administrative Region, China (CityU 123013).\\
\\
\textit{Conflict of Interest: none declared.}
\bibliographystyle{natbib}
\bibliography{ref}

\end{document}